\documentclass[10pt,conference]{IEEEtran}
\IEEEoverridecommandlockouts

\usepackage{array}
\newcolumntype{L}[1]{>{\raggedright\let\newline\\\arraybackslash\hspace{0pt}}m{#1}}
\newcolumntype{C}[1]{>{\centering\let\newline\\\arraybackslash\hspace{0pt}}m{#1}}
\newcolumntype{R}[1]{>{\raggedleft\let\newline\\\arraybackslash\hspace{0pt}}m{#1}}
\usepackage{makecell}
\usepackage{tabularx}
\usepackage{booktabs}
\usepackage{flushend}

\usepackage{cite}
\usepackage{amsmath,amssymb,amsfonts}
\usepackage{algorithmic}
\usepackage{graphicx}
\usepackage{textcomp}
\usepackage{xcolor}
\def\BibTeX{{\rm B\kern-.05em{\sc i\kern-.025em b}\kern-.08em
    T\kern-.1667em\lower.7ex\hbox{E}\kern-.125emX}}
\begin{document}

\title{The Risk-Taking Software Engineer:\\ A Framed Portrait}

\author{\IEEEauthorblockN{Lorenz Graf-Vlachy\IEEEauthorrefmark{1}\IEEEauthorrefmark{2}}
\IEEEauthorblockA{\IEEEauthorrefmark{1}\textit{Institute of Software Engineering, University of Stuttgart, Stuttgart, Germany}\\
lorenz.graf-vlachy@iste.uni-stuttgart.de}
\IEEEauthorblockA{\IEEEauthorrefmark{2}\textit{TU Dortmund University, Dortmund, Germany}\\
lorenz.graf-vlachy@tu-dortmund.de}
}

\maketitle

\begin{abstract}
\textit{Background:} Risk-taking is prevalent in a host of activities performed by software engineers on a daily basis, yet there is scant research on it. \textit{Aims and Method:} We study if software engineers’ risk-taking is affected by framing effects and by software engineers’ personality. To this end, we perform a survey experiment with 124 software engineers. \textit{Results:} We find that framing substantially affects their risk-taking. None of the ``Big Five'' personality traits are related to risk-taking in software engineers after correcting for multiple testing. \textit{Conclusions:} Software engineers and their managers must be aware of framing effects and account for them properly.
\end{abstract}

\begin{IEEEkeywords}
Risk-taking, framing, personality, five-factor model, Big Five
\end{IEEEkeywords}

\section{Introduction}

Risk-taking is prevalent in a great variety of decisions. People take risks when deciding which mating partners to choose, how to finance their homes, or which food to eat. Consequently, it is not surprising that risk-taking is one of the most extensively studied topics in a great many academic disciplines, including psychology, economics, and medicine~\cite{Yates.1992}.

Notably, risk-taking is also fundamental to many decisions software engineers make on a daily basis. These can be ``big'' decisions like choosing a software architecture or deciding which programming language to use, or ``small'' decisions on how well to document a minor change in code or whether to skip a test.
Imagine, for instance, a software engineer who has the choice between two different libraries to accomplish a given programming task. One library has precisely the needed functionality but has not seen a new release in a while and it is unclear when and if updates and patches will become available. The other library has only limited functionality but a clear roadmap for future releases. Both options have advantages and disadvantages, but choosing the first option is likely riskier than opting for the second one.

However, there is little systematic research on what determines software engineers’ risk-taking. A lot of research has focused on the management of risk, often at the project or organizational level~\cite{Masso.2022}, but there is hardly any work on the willingness of individual software engineers to take on risk. This is surprising given how consequential choices by software engineers can be and that a certain level of risk-taking has even been described as a desirable quality in software engineers~\cite{Li.2015}.

In this paper, we attempt to remedy this shortcoming by studying two especially interesting antecedents of risk-taking. For one, we consider an external factor, i.e., the ``framing'' of a decision that may influence risk-taking. For another, we study a potentially critical internal factor, i.e., the software engineer's personality. This duality of internal and external factors is particularly reasonable to consider because psychology research repeatedly demonstrated that individuals' decisions are determined both by the situation they find themselves in as well as their individual predispositions~\cite{Furr.2021}.

We thus attempt to answer the following research questions:

\begin{itemize}
    \item \emph{RQ1}: Does framing affect software engineers' risk-taking?
    \item \emph{RQ2}: Does software engineers' personality affect their risk-taking?
\end{itemize}         

\section{Theoretical Background and Related Work}

\subsection{Framing and Risk-Taking}

It has long been known that how choices are presented to individuals greatly influences the decisions they make. A particularly influential paradigm in this regard has been developed in the so-called ``heuristics and biases'' literature. Specifically, Tversky and Kahneman introduced the idea that the ``framing'' of a choice, i.e., whether it is worded in terms of potential gains or losses (while remaining logically the exact same choice), has a profound implication on respondents' level of risk-taking~\cite{Tversky.1981}. They studied different wordings (with the same expected outcome) and found that choices described as losses induce higher risk-taking than choices described as gains. These results have since been replicated in various studies~\cite{Fagley.1990, Kuhberger.1998, Steiger.2018, Druckman.2001, DeKay.2022}.

The software engineering literature includes substantial work on heuristics and biases in general.
Researchers have, for instance, found that developers are susceptible to temporal discounting~\cite{Becker.2018, Fagerholm.2019}. Other scholars found proof that developers can be substantially biased by anchoring effects~\cite{Shepperd.2018} and that selection bias leads to project overruns~\cite{Jrgensen.2013}.

Yet, there is little framing-specific research. A recent mapping study on biases in software engineering identified only three studies on framing~\cite{Mohanani.2020}. However, they either only cursorily treat the subject or they take a much looser definition of framing, allowing for substantive differences in task descriptions (e.g., labeling desired system properties as either ``requirements'' or ``ideas''). Further, a recent qualitative field-study on biases in software development in general mentioned framing. However, it lumped the specific effect of framing into a larger category of biases caused by superficial thinking, neglecting the fact that framing effects tend to persist even in situations where individuals fully reason through their choices \cite{Chattopadhyay.2020}. In another recent qualitative study of biases and architectural technical debt, which is closely related to risk-taking, framing was mentioned as a potential influence factor, although it was the least frequently mentioned one \cite{Borowa.2021}.

The most closely related work to ours is probably a study of student decision-makers who had to make requirement selection decisions. They were susceptible to a framing effect and became more risk-seeking when choosing between requirements formulated in terms of cost,
compared to when choosing between requirements formulated in terms of revenue~\cite{Fogelstrom.2009}.

\subsection{Personality and Risk-Taking}

Although there are many different personality models in the psychology literature, the currently dominant one is arguably the five-factor model~\cite{John.2021b, Feldt.2008}. As the name suggests, it comprises five personality traits, frequently also referred to as the ``Big Five''. These are openness to experience, conscientiousness, extraversion, agreeableness, and emotional stability (sometimes also referred to as its inverse, neuroticism).

Psychologists have repeatedly linked these personality traits to risk-taking, although with partially inconsistent findings. Some scholars, for instance, found that high extraversion and openness, combined with low neuroticism, agreeableness, and conscientiousness, is particularly predictive of risk-taking~\cite{Nicholson.2005}. Other researchers found extraversion and agreeableness to be the key predictors of risk-taking \cite{Joseph.2021}.

Empirical software engineering also already has a rich tradition of studying the personality of people involved in software engineering~\cite{Feldt.2008, Feldt.2010}. Scholars have, for example, used the Big Five personality framework to study the effect of developers' personality on the likelihood of pull-request acceptance~\cite{Iyer.2021}. Similarly, other studies found that committers' personality is linked to their behavior in FLOSS projects~\cite{ParumaPabon.2016}. In addition, research has found that developers higher in openness to experience make more contributions to open source software projects~\cite{Calefato.2019}. Finally, there is extant research linking personality to programming styles~\cite{Karimi.2016}.

At the same time, there is no research that we are aware of that explicitly attempts to link personality and risk-taking in a software engineering context.

\section{Empirical Setup}

\subsection{Stimulus Material and Measures}

We took inspiration from Tversky and Kahneman's original so-called ``Asian disease'' problem~\cite{Tversky.1981}, an implementation of a framing study that has been frequently used in subsequent research. To create ecological validity for our context, we adjusted the stimulus material's wording to relate it to a common software engineering problem, i.e., project delays.

Participants were randomly assigned to one of two conditions. In both conditions, participants had to make a choice between two options. The two options were substantively the same across conditions. The conditions only differed in how these options were described, or ``framed''. In the first condition, the options were framed as ``gains'', i.e., participants read about their chance of recovering time. Participants in this gain condition read the following text:

\medskip
\begin{quote}
Imagine that you are working on a software project with a deadline. You just realized that some requirements were implemented incorrectly, and you estimate that this will make you miss the deadline by 6 weeks. You think about potential remedies, and you come up with two options. You can only choose one.

\medskip
(A) If you reduce non-essential features, you will recover 2 weeks.

\medskip
(B) If you simplify the software architecture, there is a 1/3 chance that you will recover the full 6 weeks, and there is a 2/3 chance that the simplified architecture will lead to performance problems and you will not recover any time at all.

\medskip
Which option do you choose?
\end{quote}

\medskip
In the second condition, the options were described in terms of ``losses'', i.e., participants read about the delay with which they would finish the project. In this loss condition, the participants were given the following options:

\medskip
\begin{quote}

\medskip
(A) If you reduce non-essential features, you will finish with a delay of 4 weeks.

\medskip
(B) If you simplify the software architecture, there is a 1/3 chance that you will finish the project with no delay at all, and there is a 2/3 chance that the simplified architecture will lead to performance problems and you will finish with a delay of 6 weeks.

\end{quote}

\medskip

After participants made their choice, they were forwarded to further screens on which they were asked for demographic and personality information. We captured programming experience by asking for respondents' number of years of experience~\cite{Feigenspan.2012}. We employed the widely used Ten-Item Personality Measure (TIPI) to capture respondents' personality~\cite{Gosling.2003}. Since this measure has been used extensively across different populations, we have no reason to doubt its suitability to assess the personality of software engineers.

\subsection{Power Analysis and Participant Recruitment}

We performed a power analysis using G*Power 3 \cite{Faul.2007} to avoid false positives and false negatives in the analysis of framing (\emph{RQ1}) due to a potentially underpowered study. Specifically, we performed a power analysis for a $z$-test for proportions. We assume the relevant proportions of respondents choosing the risk-taking option to be 0.1 in the gain condition and 0.3 in the loss condition based on introspection and the stereotype that software developers overall might be fairly risk-averse, as well as a presumed limited strength of our stimulus material. This translates into a medium effect size of $h$ = .52~\cite{Cohen.1988}. Conservatively specifying a two-tailed test, and setting desired alpha to 0.05 and desired power to 0.80, we obtain a critical $z$-value of -1.96. Further assuming an even split of participants between conditions, this implies that a sample of 124 participants is needed. Given the number of assumptions needed for a probit (or logit) power analysis, which would be needed for our analysis of personality (\emph{RQ2}), and the limited empirical grounds we have to make them, we opted not to perform one.

To recruit participants, we obtained the contact information of all developers who made at least one commit to one of the 29 Apache open source projects that are part of the ``Technical Debt Dataset'' in version 2~\cite{Lenarduzzi.2019}.
We then identified all individuals listed as ``authors'' in the resulting data, and manually cleaned the data to remove duplicates and merge records for individuals who used different names (but the same email address) or different email addresses (but the same or an extremely similar name) for different commits. This required occasional judgment, and decisions about the identity of authors were made as conservatively as possible. In the end, we had a list of 1,555 unique individuals and one or more corresponding email addresses. To avoid excessive spam, we selected only one email address per person, preferring personal email addresses over professional email addresses to maximize the chance that the email address was still valid despite the person's contribution(s) to the projects being potentially already several years old. We invited all 1,555 developers to participate in our survey experiment (which was part of a larger data collection effort for multiple studies).

We assured the developers that their data would be treated confidentially and not be shared with third parties, and we pledged to donate US\$ 2 per completed response to the United Nations World Food Programme~\cite{Baltes.2022}. We sent two reminders to reach developers that were busy at the time of the initial mailing or who had started but not completed the survey~\cite{Baltes.2022}, including a link to an official university page confirming the authenticity of the survey because some developers responded to the initial invitation, voicing concerns about it being a scam.

In total, 165 emails bounced, allowing us to reach 1,390 developers (89.4\% deliverable emails). Of this group, 194 developers started the survey, and 124 completed it.
Our response rate was thus 8.9\%, which is in line with that of prior studies surveying developers on GitHub. Graziotin et al., for example, reported a 7\% response rate and a share of 96.6\% of deliverable emails~\cite{Graziotin.2017}.

Given that our ultimate number of participants surprisingly corresponds exactly our calculated sample size and the randomized assignment of participants to conditions lets us expect an approximately even distribution between them,
we conclude that our experimental study is sufficiently powered.

\section{Data Analysis and Results}

We first turn to the analysis for \emph{RQ1}. To study whether framing had an effect on risk-taking, we compare the share of risk-taking responses between the gain and the loss condition. To this end, we employ a two-sample test for proportions (\texttt{prtest} in Stata 17.0). Out of 63 respondents in the gain condition, 7 chose the risk-taking option. Out of 61 respondents in the loss condition, 19 chose the risk-taking option. The results of the test for proportions are shown in Table \ref{tab:framing-test-of-prop} and indicate that risk-taking is statistically significantly ($p$~$<$~0.01) higher in the loss condition. An unreported probit regression with a binary indicator of framing, as well as a two-sample Wilcoxon rank-sum test corroborate this result.

\begin{table}[ht]
\def\sym#1{\ifmmode^{#1}\else\(^{#1}\)\fi}
\small
\caption{Two-Sample Test for Proportions \label{tab:framing-test-of-prop}}
\begin{tabular}{lcccc} 
\toprule

Framing & Observations & Mean choice & $z$ & $p$ \\

\midrule
Gain & 63 & .111 &  & \\

Loss & 61 & .311 &  & \\
\midrule
Difference & & -.200 &  -2.740 & 0.006\sym{**}\\

\bottomrule
\multicolumn{5}{l}{\footnotesize Risk-averse choice coded as 0, risk-taking choice coded as 1.}\\
\multicolumn{5}{l}{\footnotesize \sym{+} \(p<0.1\), \sym{*} \(p<0.05\), \sym{**} \(p<0.01\)}\\
\end{tabular}
\end{table}

To answer \emph{RQ2}, we performed a probit regression. Aside from a binary indicator of the task framing, we included our measure of programming experience and all Big Five personality traits as independent variables. Our dependent variable was a binary indicator of whether the participant's choice was risk-taking (1) or not (0). The results are shown in Table \ref{tab:framing-personality-reg}. The indicator for loss framing is highly significant, again confirming our earlier findings. The coefficient of programming experience is not significant. More importantly, the coefficient for conscientiousness is negative and statistically significant ($p$~$<$~.05) and the coefficient for emotional stability is positive and marginally significant ($p$~$<$~.1). However, if we (despite considerable disagreement in the applied literature as to its necessity~\cite{Rothman.1990, Bender.2001, OKeefe.2003}) perform a Westfall-Young correction for multiple testing (which is more efficient that the Bonferroni method~\cite{Westfall.1993}) to limit the family-wise error rate (using Stata's \texttt{wyoung}~\cite{Jones.2019}), all coefficients for Big Five traits become insignificant (the smallest $p$-value being that for conscientiousness at .168).

\begin{table}[ht]
\def\sym#1{\ifmmode^{#1}\else\(^{#1}\)\fi}
\small
\caption{Probit Regression \label{tab:framing-personality-reg}}
\begin{tabular}{lcccc} 
\toprule

Variable & Coeff. & Std. err. & $z$ & $p$ \\

\midrule
Loss framing & .835 & .291 & 2.87 & .004\sym{**}\\
Programming experience & .141 & .108 & 1.31 & .190\\
Openness to experience & .116 & .142 & 0.81 & .415\\
Conscientiousness & -.292 & .130 & -2.24 & .025\sym{*}\\
Extraversion & -.092 & .102 & -0.91 & .364\\
Agreeableness & -.177 & .128 & -1.38 & .168\\
Emotional stability & .191 & .113 & 1.69 & .092\sym{+}\\
\midrule
Constant & .119 & .264 & -0.96 & .339\\

\bottomrule
\multicolumn{5}{l}{\footnotesize Dependent variable: Indicator of risk-aversion (0) or risk-taking (1).}\\
\multicolumn{5}{l}{\footnotesize \sym{+} \(p<0.1\), \sym{*} \(p<0.05\), \sym{**} \(p<0.01\)}\\
\end{tabular}
\end{table}

\section{Discussion}

Software engineers overall appear to be highly risk-averse. Across conditions, only 21.0\% of software engineers made a risk-taking choice despite it having the same expected outcome as the risk-averse choice. This corroborates the common stereotype of risk-averse programmers. At the same time, our results show clearly that software engineers are highly susceptible to framing effects, suggesting that the possible perception of programmers as particularly rational individuals may be misguided.

\subsection{Implications for Research}

There are several implications for research. First, we showed a framing effect for a scenario related to project delays. This raises the question for which other types of decisions or risks framing effects might exist in software engineering, and for which there might be no such effects. Similarly, our findings also raise the question if such effects are stronger or weaker for different types of roles in software development teams. In fact, one might suspect, for instance, that there could be interactive effects between task type and decision-maker role.

Second, one might wonder how to attenuate framing effects. Since the influence of framing can be considered a bias, future researchers might wish to study the effectiveness of so-called debiasing interventions in software engineers. Given that biases are mutual properties of people and tasks~\cite{Ralph.2011}, there are two avenues for debiasing. On the one hand, one might attempt to debias individuals themselves, as has for instance been proven effective with software engineers regarding the anchoring bias~\cite{Shepperd.2018}. On the other hand, one might study external influences as debiasing interventions. Prior research in other disciplines has, for example, found that strong warning messages may attenuate framing effects~\cite{Cheng.2010}.

\subsection{Implications for Practice}

\subsubsection{Developers}
The key implication for individual developers is to realize that there might be different perspectives to take on any given situation. Explicitly constructing alternative formulations of a choice might help reach more balanced decisions that are less strongly affected by framing.

\subsubsection{Managers}
Managers of software projects may want to consciously consider framing in their communication with software developers. On the one hand, this is so they do not inadvertently trigger risk-taking or risk-averse behavior in developers. They may, for instance, do so by providing multiple alternative formulations of tasks or requests. On the other hand, they might use framing purposefully as a technique to increase or decrease developers' risk-taking.
Further, project managers might wish to consider the idea of assigning roles to individual developers when important decisions are to be made. They might, for instance, ask one developer to think about a task in terms of gains and one in terms of losses. A discussion between the two might lead to the best outcome.

As the results for personality were not significant after correcting for multiple testing, we are hesitant to infer any implications, e.g., for team composition, from them.

\section{Threats to Validity}

\subsection{Construct Validity}

As one may challenge the accuracy of our measurements, we highlight that all of our measures are established and validated scales. At the same time, we recognize that the nature of short scales like the TIPI potentially introduces substantial noise into our measurement, which might also explain why our results are not significant with regard to personality.\footnote{Note that the TIPI is designed to capture all facets of the Big Five with content and criterion validity with one item each, making reliability measures like Cronbach's $\alpha$ uninformative \cite{Gosling.}. We thus do not report any.}

\subsection{Internal Validity}

While we contend that our experimental study has high internal validity, our analysis on personality may suffer from deficiencies. Critically, personality is of course not randomly assigned to participants, making it possible that we missed relevant control variables that would confound our results.

In addition, the number of participants is somewhat low for a regression analysis with as many predictors as we include. Our conclusions of no personality effects might thus also be driven by low sample size and therefore be overly conservative, even though others have also reported null findings regarding developer personality \cite{Calefato.2019}.

\subsection{External Validity}

There is a risk that our findings may not generalize to other contexts. We studied developers involved in a limited number of large open source Java projects, with a limited response rate to our survey. Our sample is thus likely not representative of all software engineers \cite{Baltes.2022}. However, we also highlight that this is possibly only a minor issue for our experimental research design, which pits one group of randomly assigned developers against another group. While these findings may thus not strictly generalize to all developers, we are nevertheless able to provide internally valid results from a sample of experienced programmers \cite{Baltes.2022}. This is of course not the case for our analysis of personality, where external validity is more substantially limited.

Additionally, one might challenge whether our experiment task has external validity. For one, although this is not typically considered very problematic~\cite{Thaler.1987}, the decision is of course hypothetical. For another, some have argued that developers make many kinds of decisions, but rarely specifically decide between two options, as they had to do in our study~\cite{Ralph.2016}.

\subsection{Reliability}

Since we provide the stimulus material and there is no human judgment involved in data analysis, our research should be highly replicable. All data to repeat the analyses of the experiment is provided in this article. As we explicitly promised all participants that their data would not be shared with third parties, we can unfortunately not release the personality data.

\section{Future Plans}

We plan to extend our work in various directions. First, we aim to collect a larger and more representative sample to replicate the study on personality effects to establish whether our null finding holds. Second, we intend to use different framing scenarios addressing different types of risky decisions software engineers may be making during their work, be it in requirements engineering, programming, testing, or other activities. Third, we strive to increase external validity by moving beyond survey experiments in favor of lab or field experiments. Specifically, we aim to study software engineering students in actual software engineering situations. Fourth, we wish to study the true interactive effects of framing and personality as well as software engineering roles to understand which kinds of software engineers are more or less susceptible to framing-induced risk-taking and which contingencies exist. Finally, we consider further extending the scope of our research by using other data collection methods such as face-to-face interviews, and by studying further influencing factors, either related to the individual developer (e.g., educational background or gender) or going beyond characteristics of the individual (e.g., organizational culture).

\section{Conclusion}

This study provides novel evidence on two types of antecedents of risk-taking in software engineers. Specifically, we show that framing has a strong influence on the risk-taking of software engineers, but we did not find reliable support for an effect of personality. We encourage future studies into the critical notion of risk-taking by software engineers.

\section*{Acknowledgment}

We thank all participants and pretesters. We acknowledge helpful comments from Daniel Graziotin and Justus Bogner.

\bibliographystyle{IEEEtran}
\bibliography{IEEEabrv,base.bib}

\end{document}